\documentclass[
  ,draft            
  ]
  {aipproc}

\layoutstyle{8x11double}

\newcommand{\ltsim}{\lower.5ex\hbox{$\; \buildrel < \over \sim \;$}}
\newcommand{\gtsim}{\lower.5ex\hbox{$\; \buildrel > \over \sim \;$}}

\newcommand{\OMC}		{\mbox{${\rm O}-{\rm C}$}}

\newcommand{\Porb}{\mbox{$P_{\rm orb}$}}

\newcommand{\Porbdot}{\mbox{$\dot P_{\rm orb}$}}

\newcommand{\exo}{\mbox{EXO\,0748$-$676}}

\begin{document}

\title{X-Ray Eclipse Timing in the LMXB EXO0748-676}

\author{Michael T. Wolff}{address={E.O. Hulburt Center for Space Research, 
Naval Research Laboratory, Washington, DC 20735}}
\author{Paul S. Ray}{address={E.O. Hulburt Center for Space Research, 
Naval Research Laboratory, Washington, DC 20735}}
\author{Kent S. Wood}{address={E.O. Hulburt Center for Space Research, 
Naval Research Laboratory, Washington, DC 20735}}

\begin{abstract}
Orbital period changes are an important diagnostic for understanding 
low mass X-ray binary (LMXB) accretion-induced angular momentum 
exchange and overall system evolution.  
We present our most recent results for the eclipse timing of the 
LMXB \exo.  
Since its discovery in 1985 it has apparently undergone three distinct 
orbital period ``epochs", each characterized by a different 
orbital period than the previous epoch.  
We outline the orbital period behavior for \exo\ over 
the past 18 years and discuss the implications of this behavior 
in light of current theoretical ideas for LMXB evolution.
\end{abstract}

\maketitle

\section{Introduction}

There are eight currently known low mass X-ray binary (LMXB) systems
that undergo full or partial eclipses.
Such systems are important for the study of evolution in
LMXBs because the eclipse edges provide timing markers that make
possible the systematic observation of orbital period changes. 
The best studied orbit in this group is that of 
\exo\ which has been X-ray active since 1985 \citep{pwgg86}.
This long timeline has allowed an unprecedented look at its orbital 
dynamics \citep[see][and references therein]{whw+02}. 
The emerging picture for the orbital period behavior,
however, is anything but the expected smooth variations in \Porb\
and \Porbdot\ based on theoretical calculations 
done to date \citep[e.g., see][and references therein]{prp02}.
Rather, the observed period changes are discontinuous 
across multiple distinct epochs, 
and large apparent changes in \Porb\ of the order seconds 
can be observed on timescales as short as one orbit.
The magnitude of the observed changes in orbital period are 
much larger than expected from LMXB evolutionary theory.  
This is likely an indication that the observed variations in 
\Porb\ are short timescale effects of angular momentum 
redistribution in the system and are masking 
the underlying long term orbit evolution.
The Rossi X-Ray Timing Explorer (RXTE) satellite has allowed
us to monitor systematically the orbit of \exo\ in an effort to
delineate and understand the process of angular momentum exchange
and its effects on the \exo\ orbit. 
In this conference paper we report our recent progress. 

\section{RXTE Observations of \exo}

\begin{figure}
\includegraphics[height=4.5in,angle=270.0]{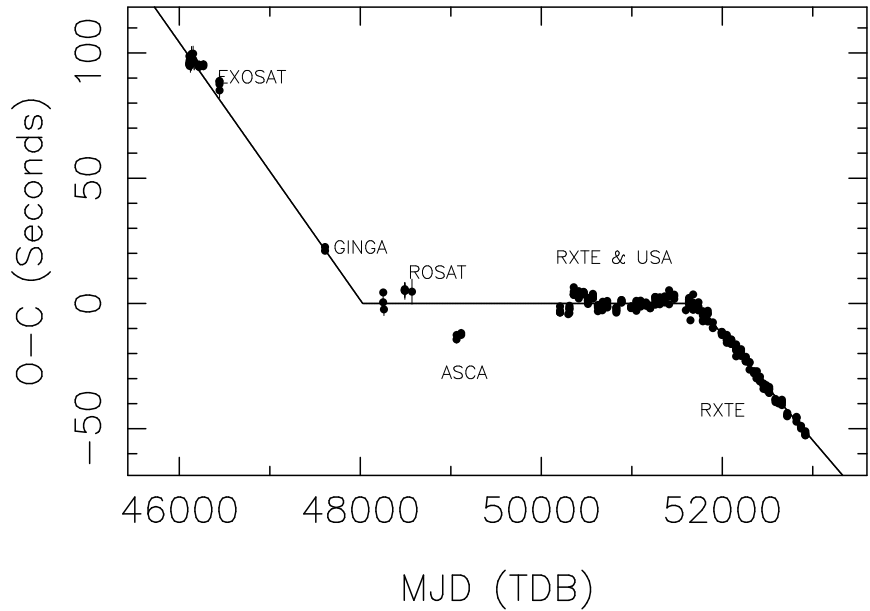}
\caption{
The \OMC\ residuals for all published mid-eclipse timings 
of \exo\ plus those additional timings we present here 
for a total of 262 mid-eclipse timings. 
All times are in Modified Julian Days corrected to the solar 
system barycenter. 
The measurement errors are as shown and except for a few cases 
are less than $\sim 1-2$ seconds. 
The solid line represents the Double-Broken Constant Period 
model for these data discussed in the text. 
The large deflection of the ASCA data away from either solution 
may be attributable to large excursions due to accumulated jitter 
as discussed in \citep{whw+02}. 
The systematic wandering of the orbit period caused by cumulative 
jitter prior to MJD 51710 is apparent. 
After this date, however, the orbit period takes on a new value 
that is only 1.4 ms larger than the apparent orbit period prior 
to MJD 48026. 
\label{fig-exoallomc}}
\end{figure}

\begin{figure}
\includegraphics[height=4.5in,angle=270.0]{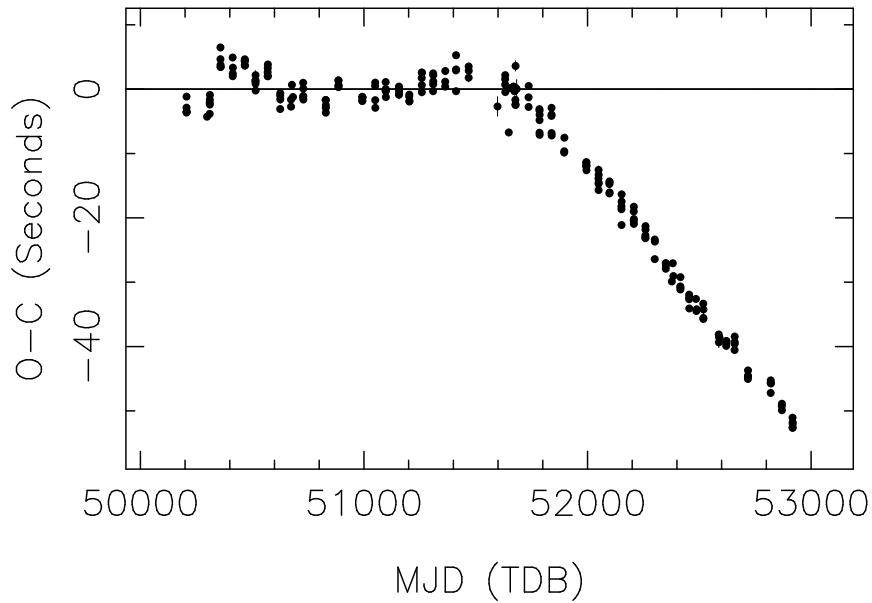}
\caption{
The \OMC\ residuals for all USA and RXTE mid-eclipse 
timings of \exo\ plotted against the middle period of the 
Double-Broken Constant Period solution in Table~\ref{tbl-exoephem}.
All times are in Modified Julian Days corrected to the solar 
system barycenter. 
The solid line represents a constant period solution for the 
\OMC\ residuals  
and is clearly unacceptable as a model for these data. 
The systematic wandering of the orbit period caused by cumulative 
jitter prior to MJD 51710 is apparent. 
After this date, however, the orbit period takes on a new value 
of $\Porb = 0.15933774511(16) \times 10^{-10}$ (days)
and the wandering about the mean period is noticeably reduced.
\label{fig-exorxteusaomc}}
\end{figure}

We now have 85 additional RXTE mid-eclipse time measurements 
over and above those analyzed in \citet{whw+02} for \exo. 
\citeauthor{whw+02} reported that the orbit period increased around 1990 
in a somewhat discontinuous manner by $\sim$7.9 ms since its discovery 
in February 1985. 
The cumulative data set of mid-eclipse timings considered in that 
paper ended in December 2000. 
Our present data set extends to observations done through October 5, 2003. 
\OMC\ (Observed - Calculated) diagrams for all published mid-eclipse 
timings of \exo\ plus the additional timings we 
present here (a total of 262 mid-eclipse time measurements) are shown 
in Figures~\ref{fig-exoallomc} and~\ref{fig-exorxteusaomc}. 

The additional eclipse measurements display a remarkable behavior.  
When the eclipses observed after December 2000 are 
included in our analysis the orbit period appears to return to 
a value close to the orbit period observed during 
the EXOSAT/GINGA epoch.
We fit a three-constant-period model (\Porb$_0$, \Porb$_1$, 
and \Porb$_2$) to these data 
in which we constrain the phase to be constant across the 
instantaneous period changes but 
let the cycles ($n_{b0}$ and $n_{b1}$) of the 
period change be free parameters (see Table~\ref{tbl-exoephem}).
The best-fit model gives a positive 
period change of 
about \Porb$_1$ $-$ \Porb$_0$ $=$ $8.19 \pm 0.06$ ms similar 
to that reported in \citep{whw+02} 
during the GINGA to early RXTE era.
However, the additional data does not continue the trend of random 
wandering about the new period. 
Instead, the apparent \exo\ orbital period undergoes 
another change near MJD 51710 (near cycle 35136) 
of \Porb$_2$ $-$ \Porb$_1$ $=$ $- 6.75 \pm 0.02$ ms. 
This brings the net change in orbit period in 
the EXOSAT$-$RXTE eras (February 1985 through October 2003) to 
only \Porb$_2$ $-$ \Porb$_0$ $=$ $1.45 \pm 0.06$ ms, significantly 
less than reported in \citep{whw+02}.
Note that the orbit period in the present epoch (\Porb$_2$)
is distinct from the period in the EXOSAT/GINGA epoch (\Porb$_0$)
according to our analysis.

\section{Cumulative Jitter in X-Ray Mid-Eclipse Timings}

The \OMC\ residuals in Figure~\ref{fig-exorxteusaomc} appear 
correlated, that is, they 
are often systematically grouped above and below the zero residual line. 
In some cases the positive or negative variations in the mid-eclipse 
residuals can persist for many months. 
We have referred to this 
variation in the apparent orbital period as ``jitter", and apparent 
changes in orbital period can be caused by accumulated intrinsic 
jitter in a constant orbital period. 
``Measurement error" is simply the random error associated with our 
measurement of the individual mid-eclipse times. 
We take ``intrinsic jitter" to mean the systematic wandering 
of the mid-eclipse timings around a smooth underlying model that 
can not be accounted for by random uncorrelated measurement errors. 
In particular, the systematic wandering of the mid-eclipse 
residuals apparent in Figure~\ref{fig-exorxteusaomc} can be 
accounted for by the accumulated intrinsic period jitter 
in individual orbit cycles. 
In order to investigate the observational measurement error 
inherent in our timing analysis and the intrinsic period jitter 
that is caused by other mechanisms (possibly mechanisms inside 
the LMXB system) we apply Maximum Likelihood Method \citep[MLM, see][]{k96} 
to estimate the parameters of a model for the orbital 
evolution that simultaneously includes an orbit period, an 
orbit period derivative, non-zero random intrinsic scatter 
(variance $\sigma_{\epsilon}^2$), and non-zero random 
measurement error (variance $\sigma_e^2$). 
We give results for \exo\ in two cases: (a) RXTE and USA data 
prior to MJD 51710, and, 
(b) the RXTE data after MJD 51710. 
A MLM model for the \exo\ RXTE and USA mid-eclipse timing data 
prior to MJD 51710 that includes a non-zero period derivative, 
intrinsic jitter and measurement error 
{\it is not} statistically preferred over a model with only 
intrinsic jitter and measurement error (a model with no intrinsic 
period jitter is strongly rejected according to the MLM analysis). 
The MLM analysis suggests that the intrinsic jitter
in the eclipse timings is characterized by $\sigma_{\epsilon} \sim 0.13$ s
with measurement error $\sigma_e \sim 1.3$ s,
similar to the results in \citep{whw+02}.
However, for the RXTE data after MJD 51710 the MLM analysis  
suggests that while the measurement error in the eclipse timings 
remains roughly constant at $\sigma_e \sim 1.2$ s, 
the intrinsic period jitter is greatly reduced 
to $\sigma_{\epsilon} \sim 0.028$ s.
Also, as was the case for the pre-MJD 51710 data, 
a model for the post-MJD 51710 timing data that 
includes a non-zero period derivative, intrinsic jitter 
and measurement error {\it is not} statistically preferred over a model 
with only intrinsic period jitter and measurement error.
Whatever process resulted in the abrupt change in the
orbital period after MJD 51710 also brought about a 
reduction in the magnitude of the intrinsic orbital period jitter
according to the MLM analysis.

\section{Discussion}

\begin{figure}
\includegraphics[height=4.5in,angle=270.0]{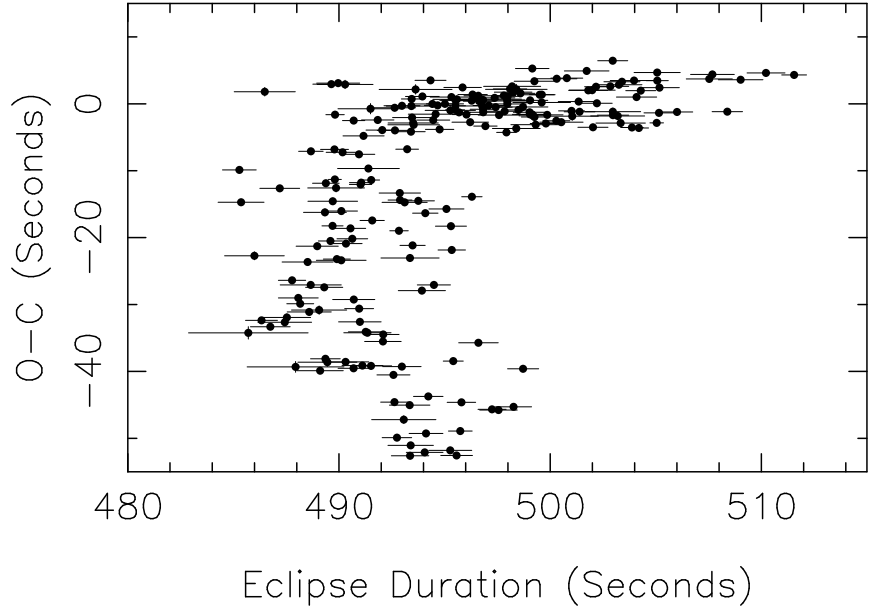}
\caption{
The durations of all RXTE observed eclipses plotted 
as a function of observed \OMC\ residual for \exo. 
The points at increasing negative \OMC\ values 
are the same eclipses observed after the dramatic change
in period near MJD 51710 in Figure~\ref{fig-exorxteusaomc}.
\label{fig-exoecldurvsomc}}
\end{figure}

The X-ray eclipse timings for \exo\ appear to show three distinct 
orbital periods in three successive epochs 
along with significant intrinsic period jitter in 
these epochs. 
Furthermore, inspection of Figure~\ref{fig-exorxteusaomc} shows 
that during the early part of the RXTE epoch (until 
approximately March 2000) the period jitter is especially prominent. 
After this date the orbit period changes again by $\sim$-6.7 ms 
nearly returning to the value characteristic of
the EXOSAT$-$GINGA epoch. 
Such abruptly changing orbit periods are observed in a number 
of Algol binaries [see the discussion in \citep{simo99} and 
references found therein] although with 
larger amplitude in ${\Delta \Porb}/{\Porb}$. 
For Algol systems ${\Delta \Porb}/{\Porb} \sim 10^{-5}$ 
whereas in \exo\ we find ${\Delta \Porb}/{\Porb} \sim 10^{-6}$.
\citep{hall91} attributed the abrupt period changes in Algol systems 
to magnetic activity in a convective secondary star inducing 
changes in its quadrupole moment and altering the orbital 
angular momentum. 
If this explanation is applicable to the period changes 
in \exo\ \citep[see also][]{hwc97} then around the 
March 2000 magnetic activity in the secondary was altered in 
some manner changing the secondary's influence on the system orbital 
angular momentum distribution. 
Before this date the orbit period jittered around a mean value
but the jitter is reduced after this date.  
This may imply that magnetic activity in the secondary also was reduced, 
perhaps as a result of magnetic cycling similar to 
the 11-year cycle in the Sun. 
Small changes in radius of the secondary are predicted by the 
magnetic activity explanation of Algol orbit period changes put
forward by \citet{ap87} due to the changing magnetic pressure support 
for the outer layers of the secondary star. 
The data in Figure~\ref{fig-exoecldurvsomc}
suggest that the radius of the
occulting star decreased across the MJD 51710 boundary. 
Totality duration is $\sim$7 seconds shorter after MJD 51710 
than before MJD 51710.
However, a spherically symmetric decrease of 1-2\% in
the radius of the secondary star is unlikely and other 
non-spherical changes in the radius
of the secondary must be considered as well.

The measurements shown in Figures~\ref{fig-exoallomc},~\ref{fig-exorxteusaomc}, 
and~\ref{fig-exoecldurvsomc}
give a magnitude and direction for the abrupt orbital period changes
and the magnitude and direction of the abrupt change in the
duration of X-ray totality.
Any attempt to understand these data will have to pose some
sort of almost instantaneous change in the binary
system parameters with an accompanying redistribution 
of the orbital and spin angular momenta rather than a slow 
change in the system because of the effects of mass exchange.

\begin{theacknowledgments}
This research is supported by the Office of Naval Research,
the NASA Astrophysical Data Program, and the 
NASA RXTE Guest Observer Program.
\end{theacknowledgments}


\begin{thebibliography}{8}
\expandafter\ifx\csname natexlab\endcsname\relax\def\natexlab#1{#1}\fi
\providecommand{\enquote}[1]{``#1''}
\expandafter\ifx\csname url\endcsname\relax
  \def\url#1{\texttt{#1}}\fi
\expandafter\ifx\csname urlprefix\endcsname\relax\def\urlprefix{URL }\fi

\bibitem[Parmar et~al.(1986)]{pwgg86}
Parmar, A.~N., White, N.~E., Giommi, P., and Gottwald, M., \emph{ApJ},
  \textbf{308}, 199 (1986).

\bibitem[Wolff et~al.(2002)]{whw+02}
Wolff, M.~T., Hertz, P.~L., Wood, K.~S., Ray, P.~S., and Bandyopadhyay, R.~M.,
  \emph{ApJ}, \textbf{575}, 384--396 (2002).

\bibitem[Podsiadlowski et~al.(2002)]{prp02}
Podsiadlowski, P., Rappaport, S., and Pfahl, E., \emph{ApJ}, \textbf{565}, 1107
  (2002).

\bibitem[Koen(1996)]{k96}
Koen, C., \emph{MNRAS}, \textbf{283}, 471 (1996).

\bibitem[Simon(1999)]{simo99}
Simon, V., \emph{A\&AS}, \textbf{134}, 1--19 (1999).

\bibitem[Hall(1991)]{hall91}
Hall, D.~S., \emph{ApJ}, \textbf{380}, L85--L87 (1991).

\bibitem[Hertz et~al.(1997)]{hwc97}
Hertz, P., Wood, K.~S., and Cominsky, L.~R., \emph{ApJ}, \textbf{486}, 1000
  (1997).

\bibitem[{Applegate} and {Patterson}(1987)]{ap87}
{Applegate}, J.~H., and {Patterson}, J., \emph{ApJ}, \textbf{322}, L99--L102
  (1987).

\end{thebibliography}

\begin{table}
\begin{tabular}{lcc}
\hline
\tablehead{1}{l}{b}{Parameter}
&\tablehead{1}{c}{b}{}
&\tablehead{1}{c}{b}{Value}\\
\hline
\multicolumn{3}{@{}c}{\bfseries All Eclipse Timing Data}\\
\hline
$T_0$ (MJD/TDB)		&$=$	&$46111.0752010(42)$\\
$\Porb_{0}$ (day)	&$=$	&$0.15933772838(66)$\\
$n_{b0}$ (cycle)	&$=$	&$12019.5(63.8)$\\
$\Porb_{1}$ (day)	&$=$	&$0.15933782322(10)$\\
$n_{b1}$ (cycle)	&$=$	&$35136.4(11.4)$\\
$\Porb_{2}$ (day)	&$=$	&$0.15933774511(16)$\\
$\chi^{2}$(dof)		&$=$	&$46.1 ( 256 )$\\
\hline
\end{tabular}
\caption{Orbital Ephemerides of \exo}
\label{tbl-exoephem}
\end{table}

\end{document}